\documentclass[a4paper,fleqn]{cas-dc}

\usepackage[numbers]{natbib}
\usepackage[noend]{algpseudocode}
\usepackage{algorithmicx,algorithm}
\usepackage{amsmath}
\usepackage{amsthm}
\usepackage{enumitem}
\usepackage{multirow}
\usepackage{makecell}
\usepackage{wrapfig} 
\usepackage{caption}
\usepackage{fontenc}
\usepackage{amssymb}
\usepackage{tabularx}
\usepackage{subcaption}

\def\tsc#1{\csdef{#1}{\textsc{\lowercase{#1}}\xspace}}
\tsc{WGM}
\tsc{QE}

\begin{document}
\let\WriteBookmarks\relax
\def\floatpagepagefraction{1}
\def\textpagefraction{.001}

\shorttitle{Items Proxy Bridging}    

\shortauthors{Zhang et~al.}  

\title [mode = title]{Items Proxy Bridging: Enabling Frictionless Critiquing in Knowledge Graph Recommendations}

\tnotemark[1] 

\author[1]{Huanyu Zhang}[type=editor,
    orcid=0000-0003-1991-565X]
\fnmark[1]
\ead{ccnuzhy@mails.ccnu.edu.cn}

\author[1]{Xiaoxuan Shen}
\ead{shenxiaoxuan@ccnu.edu.cn}

\author[1]{Yu Lei}
\ead{leiyu@mails.ccnu.edu.cn}

\author[1]{Baolin Yi}
\cormark[1]
\ead{epower@mail.ccnu.edu.cn}

\author[1]{Jianfang Liu}
\ead{jianfangliu@mails.ccnu.edu.cn}

\author[1]{Yinao xie}
\ead{xieyinao@mails.ccnu.edu.cn}

\affiliation[1]{organization={Faculty of Artificial Intelligence in Education, Central China Normal University},
            city={Wuhan},
            postcode={430079}, 
            state={Hubei},
            country={China}}

\cortext[1]{Corresponding author} 

%

















\begin{abstract}
  Modern recommender systems place great inclination towards facilitating user experience, 
  as more applications enabling users to critique and then refine recommendations immediately. 
  Considering the real-time requirements, critique-able recommender systems typically straight modify the model 
  parameters and update the recommend list through analyzing the user critiquing keyphrases in the inference phase. 
  Current critiquing methods require first constructing a specially designated model which establish direct correlations 
  between users and keyphrases during the training phase allowing for innovative recommendations upon the critiquing, 
  restricting the applicable scenarios. 
  Additionally, all these approaches ignore the catastrophic forgetting problem, where the cumulative changes 
  in parameters during continuous multi-step critiquing may lead to a collapse in model performance. 
  Thus, We conceptualize a proxy bridging users and keyphrases, proposing a streamlined 
  yet potent \emph{\underline{I}tems \underline{P}roxy \underline{G}eneric \underline{C}ritiquing Framework} (IPGC) framework, 
  which can serve as a universal plugin for most 
  knowledge graph recommender models based on collaborative filtering (CF) strategies. 
  IPGC provides a new paradigm for frictionless integration of critique mechanisms to enable 
  iterative recommendation refinement in mainstream recommendation scenarios. 
  IPGC describes the items proxy mechanism for transforming the critiquing optimization objective of user-keyphrase 
  pairs into user-item pairs, adapting it for general CF recommender models without 
  the necessity of specifically designed user-keyphrase correlation module. Furthermore, an anti-forgetting regularizer 
  is introduced in order to efficiently mitigate the catastrophic forgetting problem of the model as a prior for 
  critiquing optimization. Extensive experimental results indicate the IPGC outperforms the state-of-the-art critiquing methods 
  under various types of critiques. Meanwhile, the results show that IPGC can effectively operate on original model with different 
  architectures and alleviate the catastrophic forgetting problem significantly.  
\end{abstract}


\begin{highlights}
  \item A new generic critiquing framework (IPGC) is proposed.
  \item IPGC can work on mostly CF knowledge graph recommendation algorithms.
  \item Novel items proxy mechanism was designed to adapt to Collaborative Filtering.
  \item IPGC effectively mitigates the catastrophic forgetting problem of multi-critiquing.
  \item IPGC outperforms baselines on sota KG recommender models and benchmark datasets.
\end{highlights}



\maketitle

\section{Introduction}\label{}

Modern recommender systems are increasingly emphasizing the user experience while exploring methods to offer better recommendations \cite{wang2022user,shu2023rah,zhang2023simulating}.
More apps are allowing users to interact with straightforward feedback and deliver immediate, effective responses with refined recommendations,
 which can significantly elevate the user experience and perceptions. For example, as shown in Figure 1(a), users can continuously 
 present suggestions to gain satisfied movies. In each interaction, the system analyzes the critiquing information and innovates 
 recommendations, while the user could choose to accept or continue with other opinions. In Figure 1(b), 
 the illustrated app proactively supplies users with various categories of keyphrases, allowing them to argue or prefer 
 these keyphrases in order to precisely capture users' preferences.

Critique-able recommender systems typically involve the training phase of the original model and the inference phase in which the 
recommendations are progressively refined by analyzing the critiquing information. Recent works mostly focus on the study of 
conversational recommender system (CRS), allowing users to critiquing both items and keyphrases \cite{luo2020latent,wu2019deep,luo2020deep,li2020ranking,yang2021bayesian}. Considering that keyphrases 
can deliver more explicit and detailed information, researchers place more attention on the keyphrases. While these methods 
have proven to be capable of adapting recommendations from critiquing in CRS scenarios, there are two concerns remain to be addressed. 
Firstly, the previous researches have all designed specifically models that can directly model the correlation between users 
and keyphrases in the training phase for the benefit of accepting critiquing in the inference phase. These approaches, though 
enabling the model to handle critiquing, restrict the available scenarios and exhibit weak initial recommendation performance. Secondly, a consistent phenomenon 
present in the experimental results of these previous works is that the tail of multi-step critiquing always shows a trend of drastic performance 
decline, implying that they are not capable of handling the demands of more round interactions. This limitation stems from their 
shared neglect of inference-phase parameter drift in latent variables, where successive modifications 
induce progressive erosion of critical feature representations encoded during initial model training, 
ultimately manifesting as catastrophic forgetting \cite{zenke2017continual,aljundi2018memory,zhai2023investigating}.

\begin{figure}[t]
  \centering
  \includegraphics[width=\linewidth]{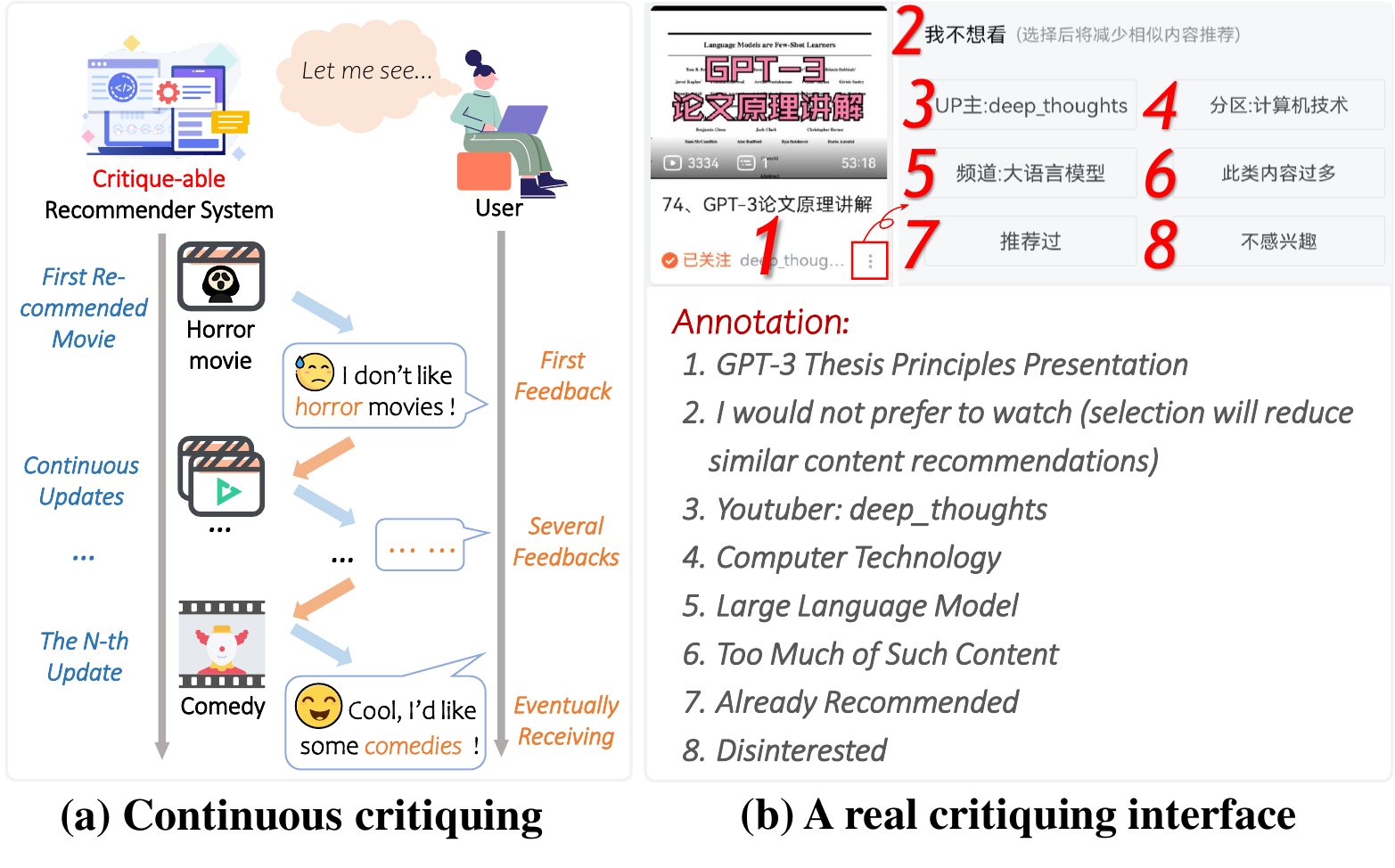}
  \caption{(a) Illustration of continuous critiquing interactions (b) Alternative critiquing keyphrases available from the video app.}
  \label{Figure1}
\end{figure}

Hence, the integration of critiquing in generalized applications via knowledge graphs shows measurable potential to relieve specific 
limitations characterizing current methods. Recent evidence \cite{wang2018ripplenet,wang2019kgat,wang2021Kgin} has indicated that knowledge graph-enhanced recommendation 
models are very powerful. Furthermore, knowledge graph (KG) could help critique-able recommender system to better understand user 
critiquing as a data structure containing rich knowledge and semantic information \cite{ma2019explainable,wang2022reasoning,tang2023logicrec}. Nevertheless constructing a universal 
critique-able model based on the KG recommendation method can be difficult. Most of these efforts are designed on collaborative 
filtering (CF) strategy, only modeling the direct users' preferences on items, and without structuring the straightforward 
correlation between users and keyphrases. Thus, we need to design a method that is generalizable among CF recommendation models \cite{xia2023graph,he2017neural} 
to represent the relationships between users and keyphrases. Additionally, the critical yet underexplored phenomenon of catastrophic 
forgetting in sequential critique scenarios poses significant risks, where progressive parameter drift during multi-step 
refinement processes may induce cascading performance degradation in recommendation systems. These observations 
necessitate the development of knowledge preservation mechanisms that maintain foundational knowledge representations 
throughout iterative critiquing cycles to mitigate catastrophic knowledge loss.

Consequently, we introduce a bayesian \emph{\underline{I}tems \underline{P}roxy \underline{G}eneric \underline{C}ritiquing Framework} (IPGC), 
that can operate as a plugin for most current major KG recommender systems relying on CF strategy, responding to user feedback in 
real-time and having the capability of anti-forgetting. For constructing the intrinsically absent user-keyphrase associations in 
generally CF models, we solve for these keyphrases by identifying proxy bridging to achieve the delivery of critiquing information, 
indirectly structure the correlations between users and keyphrases. This methodology enables zero-interference critiquing 
integration while preserving baseline recommendation integrity. Obviously, there definitely exists two optimization objectives 
in KG recommender model which are user's preference for items and the relationships between items and keyphrases, respectively. 
Using items as the proxy is thus a very rational selection. Moreover, the employment of the items proxy permits us to concentrate 
only on the direct connections between users and items during the inference phase, making it naturally adapted to the original 
CF model. Since different KG recommenders 
are mostly different in the manner of training KG structure, a generic method to match the relevance among items proxy and keyphrase 
is quite challenging to realize.We choose to leverage the most intuitive approach by treating all the items in KG linked to the 
keyphrase as its proxy and designing a multi-hop sampling protocol to ensure unbiased preference estimation across 
users' preferences for keyphrases. In addition, in IPGC we constructed a gradient-based anti-forgotten regularizer that labels essential 
parameters in user embedding prior and penalizes the changes of these key variables to ensure maximum prevention of the loss of 
critical information learned from the original model, thus alleviating the forgetting problem.
In general, our contributions are presented as follows:
\begin{itemize}[leftmargin=*]
\item We propose a generic IPGC framework that indirectly structures the correlation between user and keyphrase through items 
proxy bridging, which can frictionless effectively refine recommendations based on user critiquing. 
To the best of our knowledge, IPGC is the first critiquing framework with generalizability.
\item We use a gradient-based anti-forgotten regularizer which can effectively prevent the model from forgetting important characteristics in the original model during the inference phase.
\item We conducted extensive experiments on two real-world datasets with sota KG recommendation algorithms KGAT, KGIN, and DiffKG, the results demonstrate the effectiveness and generalizability of our IPGC framework.
\end{itemize}

\section{Related Work}
\noindent\textbf{Knowledge Graph Recommendation} has emerged as one of the most popular and well-performing models in the field of recommender 
systems, as they can enhance recommendations by utilizing the large amount of knowledge in KG. Early approaches like CKE \cite{zhang2016cke} and DKN \cite{wang2018dkn} 
leveraged pre-training on KG to strengthen item embeddings for improved recommendations \cite{wang2018shine}. In recent years, end-to-end models 
based on graph convolution \cite{wang2019Kgcn,wang2019kgat,wang2021Kgin} or graph attention networks \cite{wang2018ripplenet,wang2019Kgnn,wang2020ckan,dai2022personalized,zhang2023kgan} have proven to be more effective in mining information from KG, 
which are the which are state-of-the-art solutions that exhibit strong competitiveness. In addition, multimodal knowledge graph 
have been attempted to be applied \cite{sun2020multimodel}, and there are also methods that try to reduce the noise from higher-order information \cite{li2023dual}. 
Some methods \cite{wang2019multi} consider the coupling between KG representation task and the recommendation task or incorporate collaborative 
information \cite{wang2020ckan}, or try to mine the information more valuable to the recommendation task in KG. More recent schemes have 
incorporated graph contrastive learning on the basis of GNN to reduce the potential noise \cite{zou2022MCCLK,yang2022kgcl,zou2022kgic,yang2023knowledge,jiang2024diffkg,gong2024personalized}, yielding better recommendation performance. 
Overall, the objectives of researchers mainly focus on mining semantic and higher-order information in kg more efficiently and 
accurately and better integrating knowledge graph representing with collaborative filtering recommendations. While KG 
recommendations are highly competitive and can provide users with explanations, the overhead of time and space makes it challenging 
to update recommendations immediately with user critiquing.

\noindent\textbf{Critique-able Recommendation} differs from traditional recommendation systems, which only need to model user preferences for 
items, whereas it also requires modeling the correlation between users and keyphrases \cite{nema2021disentangling, yang2021bayesian,shen2022distributional,antognini2022positive,antognini2020interacting,antognini2021fast}. Recent work primarily focused on conversational 
recommender systems, gradually adapting recommendations to the practical demands of the user through iterative interactions between 
the user and the conversational system \cite{zhou2020improving,friedman2023leveraging,huang2023recommender,zhang2023variational}. However, most of these approaches, in order to model the association between users and 
keyphrases, are implemented within specially designed Variational Autoencoder (VAE) frameworks \cite{li2017collaborative,shen2019deep,shenbin2020recvae}. They achieve critiquing by 
redesigning the VAE encoder to achieve the input of directly correlating users with keyphrases into the latent space. In continuous 
critiquing task, they modify the potential parameters directly \cite{luo2020deep,nema2021disentangling} or using 
Bayesian methods \cite{yang2021bayesian,shen2022distributional} to update the user embedding. These methods can effectively incorporate critiquing to 
decode the refined recommendations. Although these methods indeed demonstrate the ability to tune recommendations from user feedback, 
they underutilize critiquing information. Furthermore, the adapted VAE models cannot be applied to traditional common recommendation 
scenarios nor handle large volumes of data; they are essentially and specifically designed for critique tasks. Additionally, to the 
best of our knowledge, BCIE \cite{toroghi2023bayesian} is the first attempt to combine KG with critiquing in CRS. It can effectively utilize the knowledge 
in KG, illustrating the potential of the KG for critiquing. Nonetheless, it remains an inherently recommender method dedicated to 
critiquing and is incapable of being employed in traditional list-based recommendation scenarios or state-of-the-art methods.

In conclusion, KG recommender algorithms have become one of the mainstream methods due to their performance advantages. 
And the utilization of KG enables better prehension of critiquing keyphrases. Therefore, we aim to establish a 
universal critique-able recommender paradigm based on KG recommendation models. This paradigm will seamlessly adapt to mostly 
CF methods in the KG recommendation scenario, patching the shortcomings of previous approaches and enabling rapid and efficient 
adaptation to user feedback. Not only overcome many of the shortcomings of previous approaches, but also integrates seamlessly with 
state-of-the-art methodologies.

\section{METHODOLOGY}
\subsection{Problem Formulation}
We first introduce the knowledge graph recommendation and formulate the critiquing task.

\textbf{Knowledge Graph Recommendation.} Given a user set ${\mathcal U}$, an item set ${\mathcal V}$, and a knowledge graph ${\mathcal G}$ containing a large number of 
semantic keyphrases ${\mathcal K}$ with the complex relationships between them. The objective of a knowledge graph recommender system 
is to predict the items that users probably interact with in the future, according to their historical interactions 
${\mathcal{D}_T} = \left\{ {\left( {u,v} \right)|u \in \mathcal{U},v \in \mathcal{V}} \right\}$ 
and the high-order information provided by KG. Additionally, it can provide recommendation explanations ${\mathcal{K}_e}$ for 
users from KG.

\textbf{Critiquing Task Formulation.} In a critique-able recommender system, users 
can also express their recognition of the recommendations, 
they can readily accept or critique the explanation keyphrases $k \in {\mathcal{K}_e}$ to make the 
results more in line with their perceptions. Specifically, given the user embedding ${\bf{u}}$, item embedding ${\bf{v}}$ generated 
by the original KG model, we aim to achieve a universal iterative critiquing mode for general KG recommender systems.
It allows the user to repeatedly interact with the system and generates an updated user embedding ${\bf{u}}$ with 
the user critiquing data ${{\mathcal D}_c} = \left\{ {\left( {u,k} \right)|u \in {\mathcal U},k \in {\mathcal K}}\right\}$ to 
renew scores for the candidate items and re-recommend top-k items.

\subsection{General KG Recommender Model}
In this section, we will introduce the basic principle of KG recommendation model, which is shown on the left side of Figure 2. 
Related work summarizes various types of KG recommendation models, which share the common idea of leveraging semantic and higher-order 
information among entities in the knowledge graph to explore potential interactions between users and items. These methods essentially 
optimize the representations of users, items, and keyphrases given a training set and knowledge graph. We represent the optimization 
objectives of the original knowledge graph model as follows:
\begin{equation}
  {{\mathcal L}_O} = \log p\left( {U,V,K|{{\mathcal D}_{T}},{\mathcal G}} \right)
\end{equation}
where $U$, $V$, and $K$ stand for user embeddings, item embeddings, and keyphrase embeddings, respectively. 
Specifically, these methods generally have two separate optimization objectives, namely optimizing user preferences for items:
\begin{equation}
  {{\mathcal L}_{REC}} = \sum\limits_{\left( {u,v} \right) \in {{\mathcal D}_{T}}} {\sum\limits_{\left( {u,v'} \right) \notin {{\mathcal D}_{T}}} {\log {{\mathcal F}_{rec}}\left( {{{\hat y}_{uv}} - {{\hat y}_{uv'}}} \right)} }
\end{equation}
and aggregating high-order information from the knowledge graph, which means optimizing the structural information 
between items and keyphrases:
\begin{equation}
  {{\mathcal L}_{KG}} = \sum\limits_{\left( {v,k} \right) \in {\mathcal G}} {{{\mathcal F}_{kg}}\left( {v,k} \right)}
\end{equation}
Besides, some methods have introduced other more complicated optimization objectives to achieve better recommendation 
performance with some results. Still, they revolve around the two optimization objectives mentioned above.

\begin{figure*}[t]
  \centering
  \includegraphics[width=.9\linewidth]{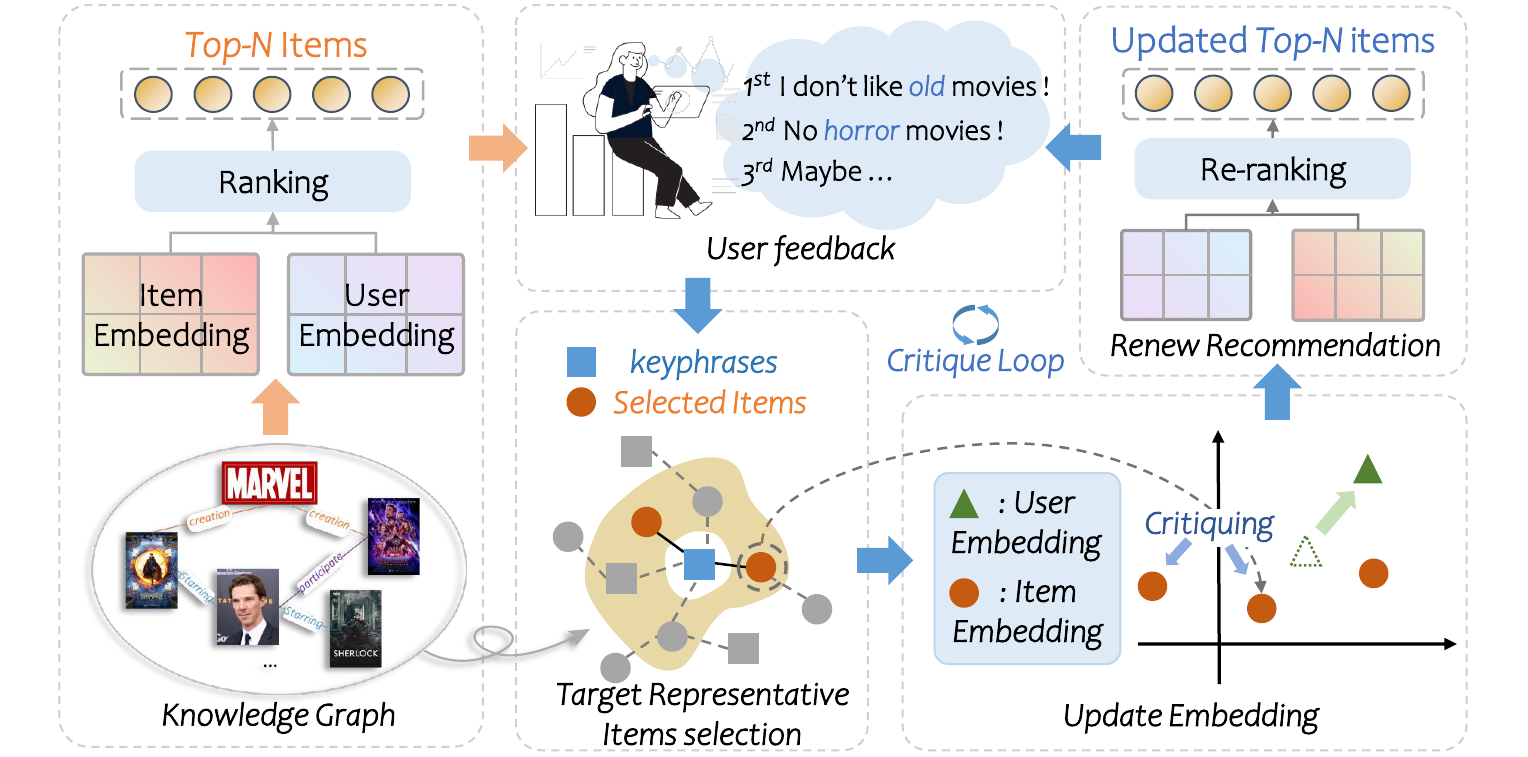}
  \caption{Illustration of the proposed IPGC, which presents the whole procedure of how the framework functions in the KG recommender system to allow critiquing for users.}
  \label{Figure2}
  \vspace{-1em}
\end{figure*}

\subsection{IPGC Framework}
The proposed IPGC targets to refine the recommendations depending on the user critiquing ${{\mathcal D}_{c}}$. In fact, we merely 
update the user embedding and keep the item embedding and keyphrase embedding invariant. The updated user embedding ${{U^ * }}$ 
can be obtained from $p\left( {{U^ * }|{{\mathcal D}_{c}},V,K} \right)$, and according to the Bayesian formula we will have:
\begin{equation}
  p\left( {{U^ * }|{{\mathcal D}_{c}},V,K} \right) \propto \underbrace {p\left( {{{\mathcal D}_{c}}|{U^*},V,K} \right)}_{critiquing\;likelihood} \times \underbrace {p\left( {{U^*}} \right)}_{user\;prior}
\end{equation}
Apply the log function to convert Eq.4 from multiplication to addition:
\begin{equation}
  \log p\left( {{U^ * }|{{\mathcal D}_{c}},V,K} \right) \propto \log p\left( {{{\mathcal D}_{c}}|{U^*},V,K} \right) + \log p\left( {{U^*}} \right)
\end{equation}
More specifically, the goal is to optimize the user embedding posterior ${{U^ * }}$ by calculating the user potential preference for the 
keyphrase, and to preserve the user embedding prior learned from the original model as much as possible. 
According to Maximum A Posteriori Estimation (MAP), the optimization objective of IPGC is to maximize the user embedding posterior estimate:
\begin{equation}
  \begin{aligned}
    {U^*} &= \mathop {\arg \max }\limits_{{U^*}} p\left( {{U^ * }|{{\mathcal D}_{c}},V,K} \right) \\
      &= \mathop {\arg \max }\limits_{{U^*}} p\left( {{{\mathcal D}_{c}}|{U^*},V,K} \right) \times p\left( {{U^*}} \right)\\
     &= \mathop {\arg \max }\limits_{{U^*}} \left( {\log p\left( {{{\mathcal D}_{c}}|{U^*},V,K} \right) + \log p\left( {{U^*}} \right)} \right)
    \end{aligned}
\end{equation}

Firstly, for the critiquing likelihood, it can be expressed as the user preference for the keyphrase:
\begin{equation}
  p\left( {{{\mathcal D}_{c}}|{U^*},V,K} \right) = \prod\limits_{\left\langle {{u_i},{k_j}} \right\rangle  \in {{\mathcal D}_{c}}} {\left( {p\left( {\left\langle {{u_i},{k_j}} \right\rangle } \right)} \right)}
\end{equation}
where ${p\left( {\left\langle {{u_i},{k_j}} \right\rangle } \right)}$ denotes the preference probability of user ${{u_i}}$ for 
keyphrase ${{k_j}}$, in this paper which will be represented by the BPR score \cite{rendle2012bpr}. Unfortunately, directly 
addressing ${p\left( {\left\langle {{u_i},{k_j}} \right\rangle } \right)}$ is challenging. 
This is because, in KG recommender model, there is no direct correlation established between users and keyphrases, making 
it impossible to calculate the BPR score between user and keyphrase through feature cross-methods.

Although it is difficult to compute ${p\left( {\left\langle {{u_i},{k_j}} \right\rangle } \right)}$ straight forward, 
the original model has already learned the preferences feature between users and items as well as the structural information 
between items and keyphrases in KG through ${{\mathcal L}_{REC}}$ and ${{\mathcal L}_{KG}}$, respectively. Considering the item 
proxy as a critical node for transmitting preference information can implicitly calculate the BPR score between users and 
keyphrases. Hence, IPGC describes a universal framework for indirectly computing the preference of users for keyphrases with item 
proxy. Formally, critiquing likelihood is transformed into the product of the BPR score between user and items proxy and 
the similarity between items proxy and keyphrases. The optimization objective is then converted to a lower bound, following Jensen's inequality we have:
\begin{equation}
  \begin{aligned}
    & \log p\left( {{\mathcal{D}_{\mathcal{C}}}|{U^*},V,K} \right) \\
    &= \sum\limits_{\left\langle {{u_i},{k_j}} \right\rangle  \in {\mathcal{D}_{\mathcal{C}}}} {\log \sum\limits_{{v_s} \in V} {p\left( {\left\langle {{u_i},{v_s}} \right\rangle } \right) \cdot p\left( {{v_s}|{k_j}} \right)} } \\
    &\geqslant \sum\limits_{\left\langle {{u_i},{k_j}} \right\rangle  \in {{\mathcal{D}}_{\mathcal{C}}}} {{\mathbb{E}_{{v_s}\sim p\left( {{v_s}|{k_j}} \right)}}\log p\left( {\left\langle {{u_i},{v_s}} \right\rangle } \right)}
  \end{aligned}
\end{equation}
Therefore, the BPR score of users for critiquing keyphrases is cleverly approximated from the preference 
expectations of users for proxy items.

Through Eq.8, we can calculate the BPR score of user-keyphrases. Practically, in the procedure of constantly 
upgrading the recommendations, the user will make new critiquing in each cycle, which will be considered for updating the 
user embedding. The observed performance degradation of multi-step critiquing is probable due to the unavoidable 
loss of essential features during continuous fine-tuning of user embedding, which has been sufficiently examined in the field of continuous learning. 
Considering the demand for real-time response in critiquing, the methods of replay not only require additional computational time expense 
and storage space but also have the possibility of causing overfitting situations. Accordingly, preserving the crucial information of the original 
model by regularization methods while optimizing the user-keyphrase BPR score is more compatible with the requirements of critiquing. 
In addition to optimizing the user-keyphrases 
BPR score in Eq.5, we hope that the updated user embedding can retain the critical information from the prior.

Ultimately, the model will compute the BPR score between the user embedding posterior and the candidate items embedding 
and re-rank them to update the Top-K recommended items for users.

\subsection{Training Strategy}
The previous section introduced the philosophy of the IPGC framework. In this section, we will describe the specific training strategy.

\noindent\textbf{Target Representative Items selection.} In Eq.8 we calculate the user-keyphrases BPR scores indirectly through 
the items proxy, making it available for the backpropagation algorithm. A crucial aspect here is to figure out the association 
probability between the items proxy and keyphrase ${p\left( {{v_s}|{k_j}} \right)}$. Since the original 
model has learned rich relationships between items and keyphrases in KG, this value can be given either by similarity 
estimation or algorithms learning KG structural features in the original model \cite{ji2015knowledge,kipf2016semi,velivckovic2017graph}. However, because of their individual 
ways of learning KG representations, it is impracticable to generalize them.

Therefore, to commonly address Eq.8, we instead decided to utilize the basic structure of KG to represent the associations 
between items and keyphrases. Specifically, if a connection trail between them is observed in KG, we 
set ${p\left( {{v_s}|{k_j}} \right)}$ to 1, otherwise 0. Considering that sampling 
all items related to keyphrase in KG incurs excessive cost, we use the Monte Carlo method \cite{shapiro2003monte} for unbiased sampling approximations:
\begin{equation}
  \begin{aligned}
  & { \mathbb{E}_{{v_s}\sim p\left( {{v_s}|{k_j}} \right)}} \log p\left( {\left\langle {{u_i},{v_s}} \right\rangle } \right) \\
   & \approx \frac{1}{M}\sum\limits_{s = 1}^M {\log p\left( {\left\langle {{u_i},v_s^*} \right\rangle } \right)} ,v_s^*\sim p\left( {{v_s}|{k_j}} \right)
 \end{aligned}
\end{equation}
where $M$ is the number of samples.
Since the learned KG representations in the original model encompass some higher-order information, to minimize the loss of 
this information due to sampling items proxy, it is necessary to sample higher-order items associated with keyphrases when using 
the Monte Carlo method. We adopt a multi-hop sampling method to extract $l$ hop items related to keyphrase $k$ from the knowledge graph:
\begin{equation}
  {{\mathcal V}^*} = MultihopSample\left( {l,r,k,{\mathcal G}} \right)
\end{equation}
where we sample items at different hops proportionally, and $r$ denotes the ratio of items sampled from the $l$-th hop over the 
total $M$ sampled items. And we apply a simple random sampling strategy for items at each hop. Therefore, Eq.9 can be transformed into:
\begin{equation}
  \begin{aligned}
  &{\mathbb{E}_{{v_s}\sim p\left( {{v_s}|{k_j}} \right)}}\log p\left( {\left\langle {{u_i},{v_s}} \right\rangle } \right) \\
  &\approx \frac{1}{M}\sum\limits_{s = 1}^M {\log p\left( {\left\langle {{u_i},v_s^*} \right\rangle } \right)} ,v_s^* \in {{\mathcal V}^*}
 \end{aligned}
\end{equation}
Thus, we can tactfully substitute the BPR score of user for items proxy with the BPR score of the user against the keyphrase:
\begin{equation}
  \log p\left( {\left\langle {{u_i},v_s^*} \right\rangle } \right) = \log \sigma \left( {h - {{\bf{u}}_i}^\top {\bf{v}}_s^*} \right)
\end{equation}
where $h$ is a constant and $\sigma$ denotes the sigmoid function. And the critiquing likelihood can be expressed as:
\begin{equation}
  \begin{aligned}
  & \log p\left( {{{\mathcal D}_{c}}|{U^*},V,K} \right) \\
  &= \eta  \cdot \frac{1}{M}\sum\limits_{\left\langle {{u_i},{k_j}} \right\rangle  \in {{\mathcal D}_{c}}} {\sum\limits_{s = 1}^M {\log \sigma \left( {h - {{\bf{u}}_i}^\top {\bf{v}}_s^*} \right)} }
\end{aligned}
\end{equation}
where $\eta$=1 means k is positive critiquing and $\eta$=-1 represents negative critiquing. Alternatively, 
IPGC provides the option of involving $(u, i)$ pairs from $\mathcal D_{T}$ in updating the user embedding when user critiquing is negative:
\begin{equation}
  \begin{aligned}
    & \log p\left( {{{\mathcal D}_{c}}|{U^*},V,K} \right) \\
    & = - \frac{1}{M}\sum\limits_{\left\langle {{u_i},{k_j}} \right\rangle  \in {{\mathcal D}_{c}},\left\langle {{u_i},{v^ + }} \right\rangle  \in {{\mathcal D}_{T}}} {\sum\limits_{s = 1}^M {\log \sigma \left( {{{\bf{u}}_i}^\top {{\bf{v}}^ + } - {{\bf{u}}_i}^\top {\bf{v}}_s^*} \right)} } 
    \end{aligned}  
\end{equation}

\noindent\textbf{Continuous Critiquing.}
In section 3.3 we analyzed the problem of recommendation performance collapse in continuous critiquing and raised the prospect of 
alleviating it by inhibiting key feature change within user embedding prior to using regularization. 
At this point, our goal is to mitigate catastrophic forgetting by tokenizing critical parameters 
and penalizing the tendency towards modifying those parameters. 
We are inspired by previous approaches \cite{aljundi2018memory}, which estimate these emphasis weights approximate the 
sensitivity of the learned function in the original model. Notice that the method was designed for computer vision and 
requires adaptation for critiquing tasks; we focus only on the last layer of the original recommendation model which is to 
compute the feature-crossing of any candidate item v for each user u. Since the item embedding is fixed invariant, we only care about the sensitivity 
of the user embedding against perturbation. For a given $(u,v)$ pair, we use $\phi \left( {{u_v};{\theta _u}} \right)$ to denote 
its BPR score and ${\theta _u} = \left\{ {{u^{\left( i \right)}}} \right\}$ to denote the user embedding tensor. Then the 
perturbation $\delta  = \left\{ {{\delta _i}} \right\}$ to ${{\theta _u}}$ resulting in change in the final predicted value can be expressed as:
\begin{equation}
  \phi \left( {{u_v};{\theta _u} + \delta } \right) - \phi \left( {{u_v};{\theta _u}} \right) \approx \sum\limits_i {{g_i}\left( {{u_v}} \right)} {\delta _i}
\end{equation}
Where the value of ${g_i}\left( u \right) = \frac{{\partial \left\| {\phi \left( {{u_v};{\theta _u}} \right)} \right\|}}{{\partial \theta _u^{\left( i \right)}}}$ can 
measure the significance of the weight, indicating how much the current prediction will change with a 
perturbation to this parameter. Finally, we can obtain the importance weight of each user:
\begin{equation}
  {\Omega _u} = \frac{1}{N}\sum\limits_{v = 1}^N {\left\| {{g_i}\left( {{u_v}} \right)} \right\|}
\end{equation}
where $N$ is the total interaction number of users in the training set. Modifications to parameters with minor importance 
weights will not have a significant impact on the output, and they will not be overly constrained in further tasks. 
However, changes to parameters with higher importance weights will be restricted or preferably left unchanged, we have:
\begin{equation}
  \log p\left( {{U^*}} \right) = \sum {{\Omega _U}\left\| {U - {U^ * }} \right\|}
\end{equation}

\noindent\textbf{Update embedding.}
Finally, the maximum posterior probability of the user embedding can be transformed into minimizing the loss function. 
Therefore, for the proposed IPGC framework, we have the following loss function:
\begin{equation}
  \begin{aligned}
    \min {{\mathcal L}_{IPGC}} =&  - \log \left( {p\left( {{{\mathcal D}_{c}}|{U^*},V,K} \right) \cdot p\left( {{U^*}} \right)} \right)\\
     =&  - \eta  \cdot \frac{1}{M}\sum\limits_{\left\langle {{u_i},{k_j}} \right\rangle  \in {{\mathcal D}_{c}}} {\sum\limits_{s = 1}^M {\log \sigma \left( {h - {{\bf{u}}_i}^\top {\bf{v}}_s^*} \right)} } \\
     & - {\lambda _\Omega }\sum {{\Omega _U}\left\| {U - {U^ * }} \right\|} 
    \end{aligned}
\end{equation}
Where ${\lambda _\Omega }$ is a hyperparameter and $v_s^* \in {{\mathcal V}^*}$.

\section{EXPERIMENTS}
We present empirical results to substantiate the efficacy of our proposed IPGC. 
The experiment is intended to answer the following research questions:

\noindent\textbf{RQ1}: Does the proposed IPGC effectively refine the recommendation in terms of critiquing information? How does the result compare with the current state-of-the-art critiquing methods regarding generalizability and performance?

\noindent\textbf{RQ2}: Does the knowledge graph manifest a positive influence on finding items proxy, and does the anti-forgetting regularizer actually work?

\noindent\textbf{RQ3}: Different hyperparameter settings (e.g., number of multi-hop sampling layers, effect of total number of samples, etc.) and case study.

\subsection{Dataset Description}
We validate IPGC on two real-world datasets for movie and music in the experiments: (1) We use the Movielens-1M, a 
widely used movie benchmark dataset released by RippleNet \cite{wang2018ripplenet}; and (2) Last-FM, a music benchmark 
dataset collected from the Last.fm online music system released by KGAT \cite{wang2019kgat}. For Movielen-1M, since 
it incorporates the user's explicit scores of the movie (rating from 1 to 5), we convert it to implicit 
feedback (with setting the rating threshold to 1). As for Last-FM we follow exactly the protocol given by KGAT 
for processing and slicing the dataset in order to avoid inaccuracy. The statistical information of these datasets is 
summarized in Table 1. Keyphrases refer to the entities in the knowledge graph that are not items. We split the datasets 
into training and testing sets at 8:2. The other training settings for KGAT, KGIN and DiffKG followed the configurations provided 
in their papers. We stopped training the original model until it got the highest recommendation score.
In the evaluation phase, we use the all-ranking strategy to evaluate the recommendations, which is to rank all the items 
excluding the training set. Our codes are shared on https://github.com/StZHY/Critique/tree/master.

\begin{table}
  \caption{Statistics of MovieLens-1M and Last-FM}
  \centering
  \label{tab1}
        \begin{tabular}{c|c|c|c} 
          \toprule
           & & Movielens-1M & Last-FM \\ 
          \midrule
          \multirow{3}*{\makecell[c]{User-Item \\ Interaction}} & Users & 6036 & 23566\\
          ~ & Items & 2445 & 48123 \\
          ~ & Interactions & 376886 & 1712638\\
          \midrule
          \multirow{3}*{\makecell[c]{Knowledge \\ Graph}} & Entities & 182011 & 106389\\
          ~ & Keyphrases & 179566 & 58266\\
          ~ & Triplets & 309172 & 464567\\
          \bottomrule
     \end{tabular}
     \vspace{-1em}
\end{table}

\subsection{Experiment Settings}
We compare IPGC with CE-VAE \cite{luo2020deep}, BK-VAE \cite{yang2021bayesian}, DCE-VAE \cite{shen2022distributional} and BCIE \cite{toroghi2023bayesian} on two real-world 
datasets: the MovieLens-1M and Last-FM. During the inference 
phase, adopting the familiar settings from previous studies \cite{yang2021bayesian,shen2022distributional}, we conducted a total of 10 rounds of multistep 
critiquing experiments. In order to simulate real user feedback without exposing the practical test dataset, we 
used only negative critiquing (positive critiquing will inevitably include attributes of the items in the test set). 
By using a program independent of IPGC, we calculated the differences between the average frequency of keyphrases 
occurrences corresponding to the Top-20 items predicted by the model in each round and corresponding to the target 
items (the results of the first round were taken from the predictions of the original model). We then selected the 
top-5 keyphrases with the largest differences as the user critiquing. To evaluate the effectiveness of the critiquing model, we used 
three metrics, \emph{Recall@5}, \emph{NDCG@5}, and \emph{HR@}5, to measure the effectiveness in refining recommendations based on 
multi-step critiquing. It should be emphasized that, as mentioned 
in previous studies \cite{yang2021bayesian,shen2022distributional}, meaningful comparisons of performance trends can only be conducted if the starting scores are roughly the 
same. However, in previous work, as shown in Table 2, the base recommendation is significantly less than the KG-based approach, 
and the VAE model is not suitable for general KG recommendation task. To ensure fair comparisons, we have re-implemented their 
methodology in the KG recommender system, although that may cause potential issues. Our codes are shared on https://github.com/StZHY/IPGC/tree/master.

\begin{table*}
  \caption{Performance of different original methods.}
  \label{tab2}
  \centering
  \begin{tabular}{c|ccc|ccc}
    \toprule
    \multirow{2}*{} & \multicolumn{3}{c|}{MovieLens-1M}  & \multicolumn{3}{c}{Last-FM} \\
      ~ & Recall@5 & NDCG@5 & HR@5 & Recall@5 & NDCG@5 & HR@5 \\
      \midrule
       CE-VAE & 0.0265 & 0.0680 & 0.1701 & 0.0096 & 0.0191 & 0.0379\\
       BK-VAE& 0.0324 & 0.0793  & 0.2037 & 0.0115 & 0.0232 & 0.0462\\
       DCE-VAE& 0.0370 & 0.0842 & 0.2171 & 0.0117 & 0.0231 & 0.0463\\
       BCIE& 0.0272 & 0.0690 & 0.1559 & 0.0238 & 0.0397 & 0.0745\\
       \midrule
       KGAT& 0.0897 & 0.1967 & 0.5172 & 0.0399 & 0.0612 & 0.1789\\
       KGIN& 0.1327 & 0.2687 & 0.6368 & 0.0483 & 0.0743 & 0.1960\\
       DiffKG& 0.1275 & 0.2589 & 0.6322 & 0.0529 & 0.0843 & 0.2154\\
       \bottomrule
  \end{tabular}
  \vspace{-1em}
\end{table*}

\subsection{Baselines}
We introduce three state-of-the-art KG recommender models \textbf{KGAT} \cite{wang2019kgat}, \textbf{KGIN} \cite{wang2021Kgin} 
and \textbf{DiffKG} \cite{jiang2024diffkg} and implement IPGC on top of them. Next, we 
present baseline critiquing methods and detail how we apply these methods towards KG recommender systems:

\textbf{CE-VAE} \cite{luo2020deep} operates by directly setting the weights of the VAE latent variable corresponding to critiquing 
keyphrase embeddings to zero. In the KG recommender framework, we exclude critiquing keyphrase from user embeddings aggregation.

\textbf{BK-VAE} \cite{yang2021bayesian} uses a Bayesian algorithm to update the user embedding through modify the corresponding keyphrases 
of users. In this paper, we update the user embeddings by directly constructing a BPR loss function with the user 
embeddings and keyphrase embeddings.

\textbf{DCE-VAE} \cite{shen2022distributional} further clarified critiquing keyphrases based on BK-VAE by using a keyphrases tree to capture 
more precise meanings of user responses. Since KG itself consists of a large amount of refined knowledge, we build on BK-VAE 
by mapping the keyphrases to all direct neighboring nodes in the KG to refine the meaning of the keyphrases.

\textbf{BCIE} \cite{toroghi2023bayesian}: It models triplets (user, like, items) in the knowledge graph through knowledge representation 
methods and transmits user feedback using belief propagation methods. We traverse the set of all nodes in the KG on the relation 
paths between the user's historical items and the critiquing keyphrase nodes as the object to update recommendations.

\begin{figure*}[htbp]
	\centering
    \begin{minipage}{1\linewidth}
		\centering
		\includegraphics[width=.9\textwidth]{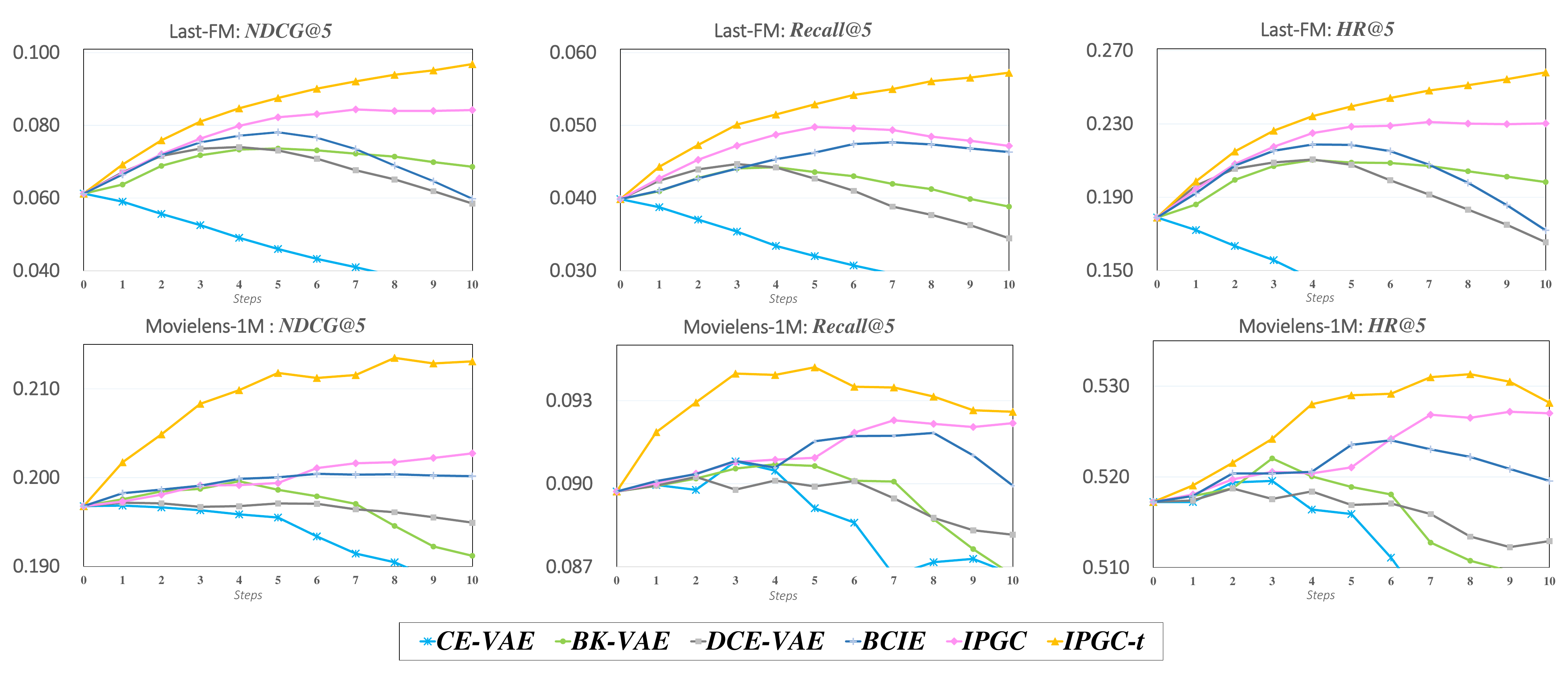}
		\caption{Comparison of IPGC and baseline methods on \emph{KGAT} for \emph{NDCG@K}, \emph{Recall@K} and \emph{HR@K}, with dataset MovieLens and Last-FM.}
		\label{fig:1} 
	\end{minipage}\\
	\vspace{0.5em} 
	\begin{minipage}[c]{1\linewidth}
		\centering
		\includegraphics[width=.9\textwidth]{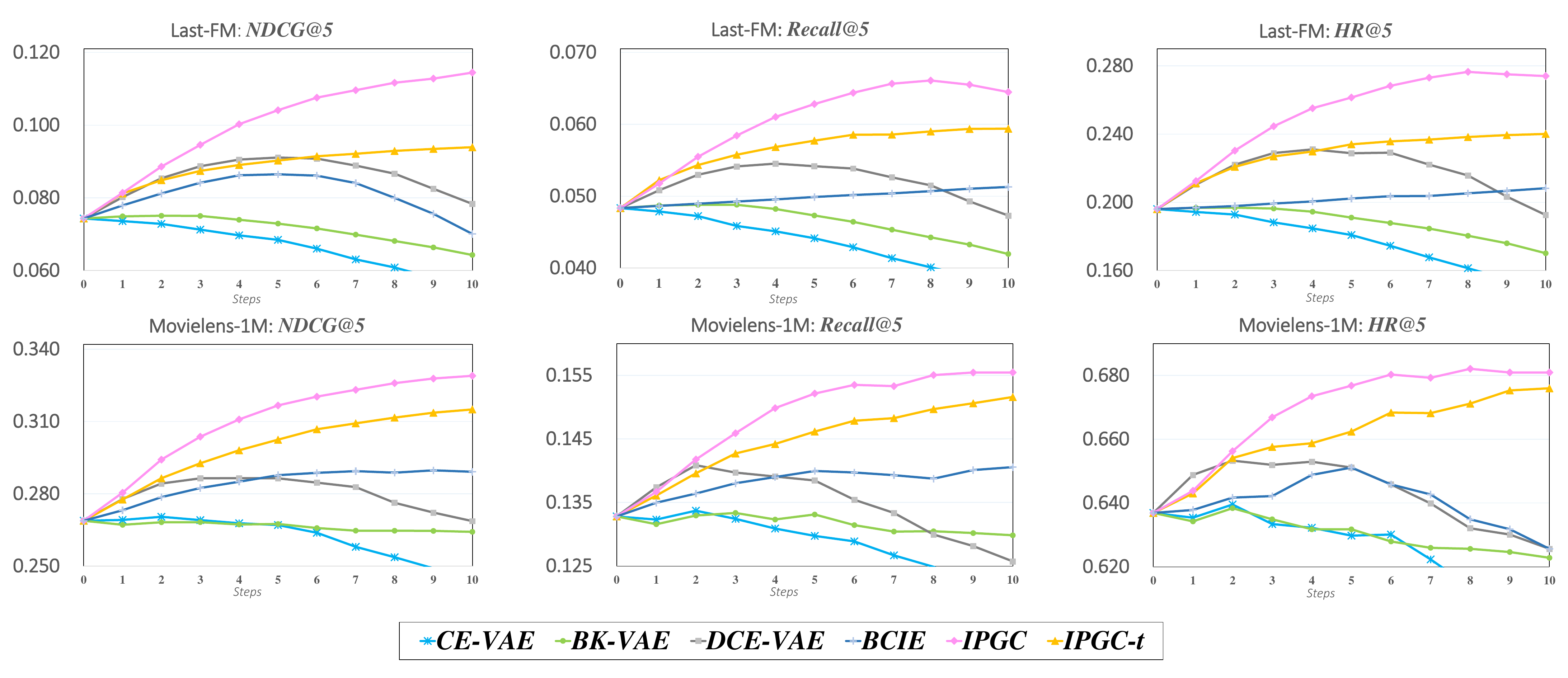}
		\caption{Comparison of IPGC and baseline methods on \emph{KGIN} for \emph{NDCG@K}, \emph{Recall@K} and \emph{HR@K}, with dataset MovieLens and Last-FM.}
		\label{fig:2} 
	\end{minipage}\\
	\vspace{0.5em} 
	\begin{minipage}{1\linewidth}
		\centering
		\includegraphics[width=.9\textwidth]{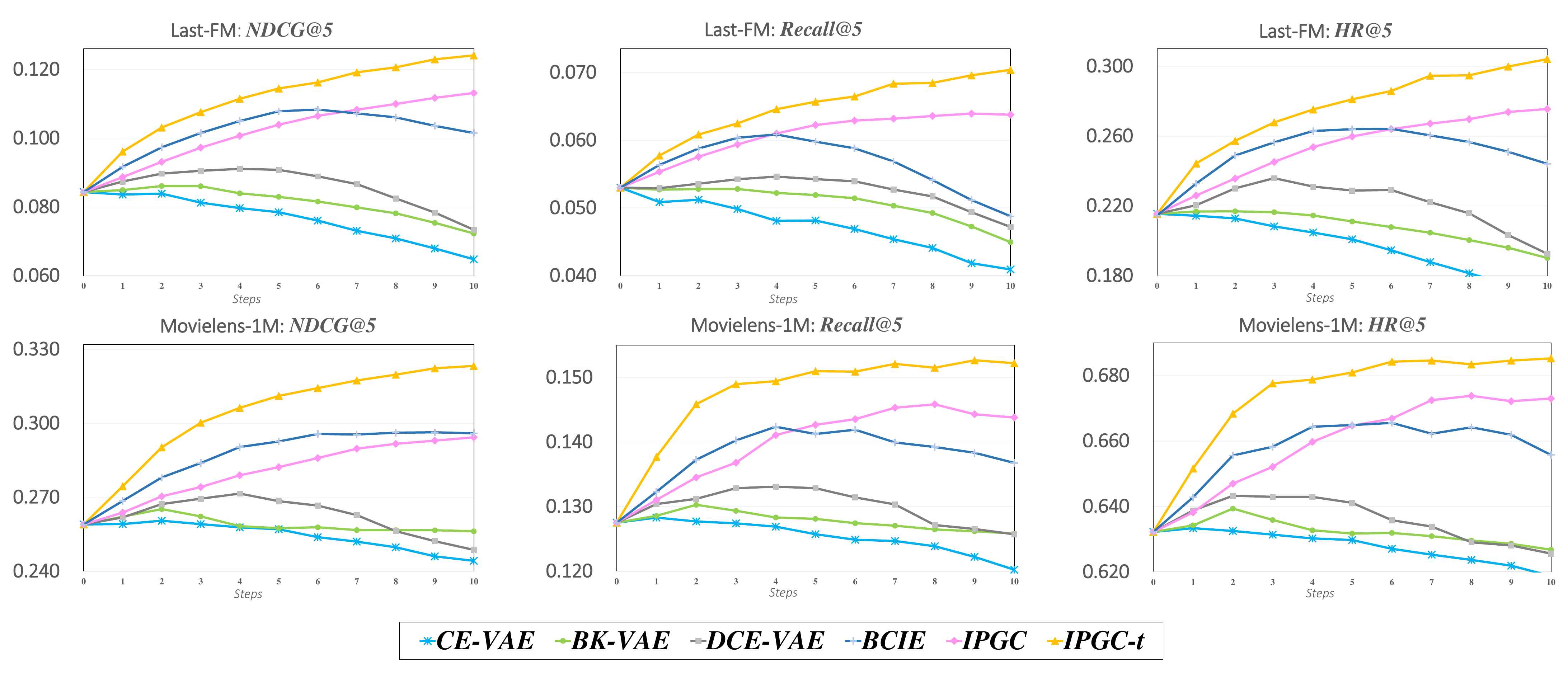}
		\caption{Comparison of IPGC and baseline methods on \emph{DiffKG} for \emph{NDCG@K}, \emph{Recall@K} and \emph{HR@K}, with dataset MovieLens and Last-FM.}
		\label{fig:3} 
	\end{minipage}
\end{figure*}

\subsection{Parameter Settings}
We implement the IPGC in PyTorch. For a fair comparison, we set the optimization method as Adam \cite{kingma2014adam}, and the learning rate 
is to $0.005$, specially fixing this value for all datasets and comparison methods. We use the grid search to explore 
the parameters of the IPGC. First the regularization parameter is set ${\lambda _\Omega }$ between $\{{10}^{ - 2}, {10}^{ - 3}, {10}^{ - 4}\}$, the hop 
number for sampling is 2, and the probability $r$ of sampling from $l$=1 is tuned in $\{1.0, 0.8, 0.5, 0.0\}$.For the total 
number $M$ of items proxy sampled for each keyphrase $k$ is chosen from $\{1, 5, 10\}$. And for the two optimization 
for \emph{IPGC}, where the one using $(u,i)$ pairs in the training set is named \emph{IPGC-t}.

\subsection{Performance Comparison (RQ1)}
The results of the multi-step critique are reported in Figure 3, 4 and 5, where the trend of the scores signifies the ability of each 
method to refine recommendations.

\begin{itemize}[leftmargin=*]
\item DiffKG, KGIN and KGAT demonstrate better performance on three VAE-based metrics and BCIE in Table 2. A better original model can serve a superior 
user experience, while increasing performance on the better model is usually more challenging.

\item IPGC exhibits significantly competitive performance compared to other critiquing methods. In a continuous 10-step 
critiquing on e.g., Last-FM, IPGC can raise the original scores of KGIN by $54\%$, $36\%$, and $40\%$ for the 
metrics \emph{NDCG@5}, \emph{Recall@5}, and \emph{HR@5}, respectively. Compared with the Sota critiquing methods, it has 
significant improvement. Although our simulation experiments represent an ideal situation, such a dramatic benefit is a 
validation of IPGC's excellent capability to refine recommendations in different methods and datasets.

\item The improvement on KGAT, KGIN and DiffKG is approximately similar, indicating that the selection of the original model does 
not impact IPGC. This suggests our model, as a plug-in, can be adapted to all general KG recommender models and can produce 
good results.

\item Compared to IPGC, IPGC-t reinforces user preferences using examples from the training data and performs better in 
KGAT and DiffKG. We conclude this result may be related to the initial model design.

\item CE-VAE, BK-VAE, DCE-VAE and BCIE perform not well. This is because they attempt to directly optimize user-keyphrase 
preference scores, which is effective for specially designed VAE models but not for a general KG-based model. Some methods 
do not even have any proactive effect. Furthermore, all these methods exhibit the typical characteristics of catastrophic 
forgetting in multi-step critiquing task, indicating that the problem we have raised is not an illusory one. Moreover, the 
weaker performance of these methods implicitly demonstrates the validity of mining items proxy to indirectly compute the BPR 
score of users for keyphrases.
\end{itemize}

\begin{figure}[]
  \centering
  \includegraphics[width=.9\linewidth]{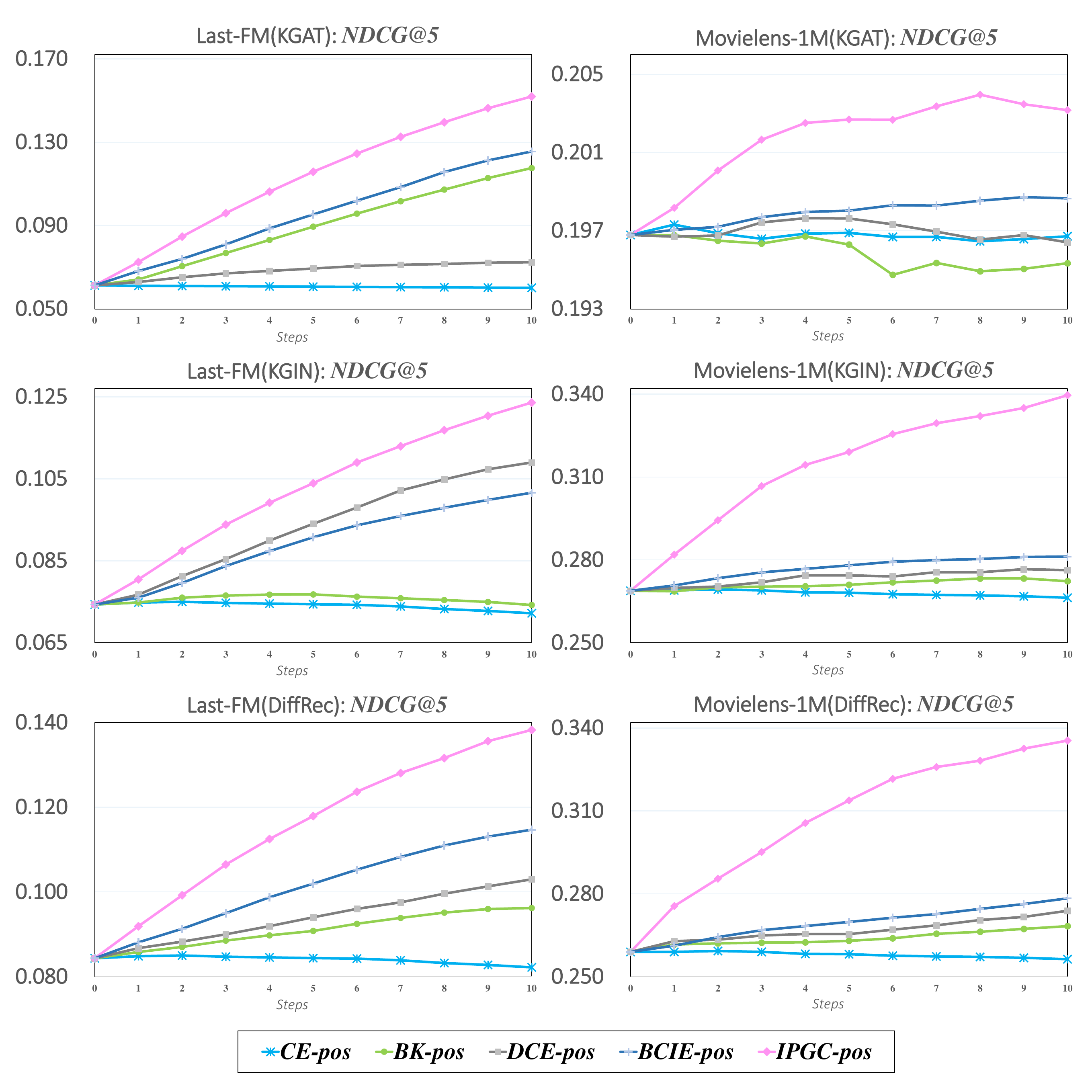}
  \caption{Comparison of IPGC and baselines based on KGAT, KGIN and DiffKG using positive critiquing at NDCG@5.}
  \label{Figure6}
\end{figure}

\subsection{Performance of Positive Critiquing (RQ1)}
Despite what we assume in the real world, most critiquing is negative because when users are clearly expressive of their 
preferences, they usually know what they want and rarely rely on recommender systems to shop for their favorite items. 
However, especially in conversational recommendation systems, a high probability of positive critiquing arises. We also conducted 
an exploration of positive critiquing, even though there will be a possibility that the test data will be leaked to the critiquing 
model. As illustrated in Figure 6, we observed that even with positive critiquing, IPGC still exhibits better performance 
and the trend does not diminish with increasing critique rounds. Noteworthy that even with positive feedback, baseline models 
still suffer catastrophic forgetting, indicating that IPGC constructs user-keyphrase correlation through items proxy is 
more reasonable.

\subsection{Analysis of the Items Proxy (RQ2)}
The effectiveness of items proxy has been demonstrated, but it is worth exploring whether the knowledge graph plays a key 
role in items proxy selection. We use Table 3 to visually illustrate the comparison between target items proxy from the 
KG and using random sampling. And it is obvious from the results that KG takes a crucial part in the search of items proxy. 
Besides, random sampling helps to improve recommendations is evident in the Movielens dataset. This is because during the 
sampling process, there is always a chance of picking items that are relevant to user critiquing, leading to some improvements. 
However, in the sparser Last-FM dataset, random sampling did not produce any significant effect.

\begin{table*}\footnotesize
  \centering
  \caption{Comparison of sampling according to KG and random sampling. The metrics are Recall@5, NDCG@5 ,and HR@5.}
  \label{tab3}
    \begin{tabularx}{\linewidth}{c|ccc|ccc|ccc|ccc}
    \toprule
    \multirow{3}*{Steps} & \multicolumn{6}{c|}{IPGC(KGIN)}  & \multicolumn{6}{c}{IPGC(KGIN)-random sampling} \\
    ~ & \multicolumn{3}{c}{Last-FM} & \multicolumn{3}{c|}{Movielens-1m} & \multicolumn{3}{c}{Last-FM} & \multicolumn{3}{c}{Movielens-1m} \\
    ~ & Recall & NDCG & HR & Recall & NDCG & HR & Recall & NDCG & HR & Recall & NDCG & HR \\
    \midrule
    0 & 0.0483 & 0.0743 & 0.196 & 0.1327 & 0.2687 & 0.6368 & 0.0483 & 0.0743 & 0.196 & 0.1327 & 0.2687 & 0.6368 \\
    1 & 0.0517 & 0.0814 & 0.2124 & 0.1367 & 0.2804 & 0.6438 & 0.0483 & 0.0742 & 0.1957 & 0.1323 & 0.2694 & 0.6368 \\
    3 & 0.0584 & 0.0946 & 0.2445 & 0.1459 & 0.3036 & 0.6668 & 0.0479 & 0.0735 & 0.1945 & 0.1334 & 0.2716 & 0.6375 \\
    6 & 0.0644 & 0.1076 & 0.2682 & 0.1535 & 0.3202 & 0.6803 & 0.0474 & 0.0726 & 0.1913 & 0.1357 & 0.275 & 0.6412 \\
    10 & 0.0655 & 0.1145 & 0.275 & 0.1554 & 0.3289 & 0.681 & 0.0467 & 0.0717 & 0.1909 & 0.1345 & 0.2752 & 0.6381 \\
    \midrule
    MaxImp & 36\%  & 54\%  & 40\%  & 17\%  & 22\%  & 6\%   & 0\%   & 0\%   & 0\%   & 2\%   & 2\%   & 0.60\% \\
    \bottomrule
    \end{tabularx}
\end{table*}

\subsection{Efficacy of Anti-forgotten Regularizer (RQ2)}
In Section 4.5, most other methods suffer from catastrophic forgetting after 2 or 3 rounds. In contrast, both 
IPGC and IPGC-t manage to avoid the inference of forgetting and show stable tuning capabilities. 
However, the stable 
refinement of IPGC and IPGC-t could also due to the fact that the learning rate is set too small. To eliminate the effect 
of this factor, we consider exploring the utility of the anti-forgetting regularizer. in Figure 7(c), where ${\lambda _\Omega }=0$ represents
the case without the regularizer, it is obvious that the regularizer does provide a great deal of assistance in preventing 
catastrophic forgetting. Without it, the model will struggle to growth after steps with the recommendation declining. 
its overall performance is weaker compared to IPGC with the regularizer. That suggests that regularizer are effective in 
inhibiting the variation of key weights and successfully preserving the user's initial preferences.

\subsection{Hyperparametric Analysis (RQ3)}
In this section we will discuss some of the hyperparameters we set in the IPGC. 

\begin{figure*}[h]

	\begin{minipage}[t]{0.32\textwidth} 
		\includegraphics[width=0.95\textwidth]{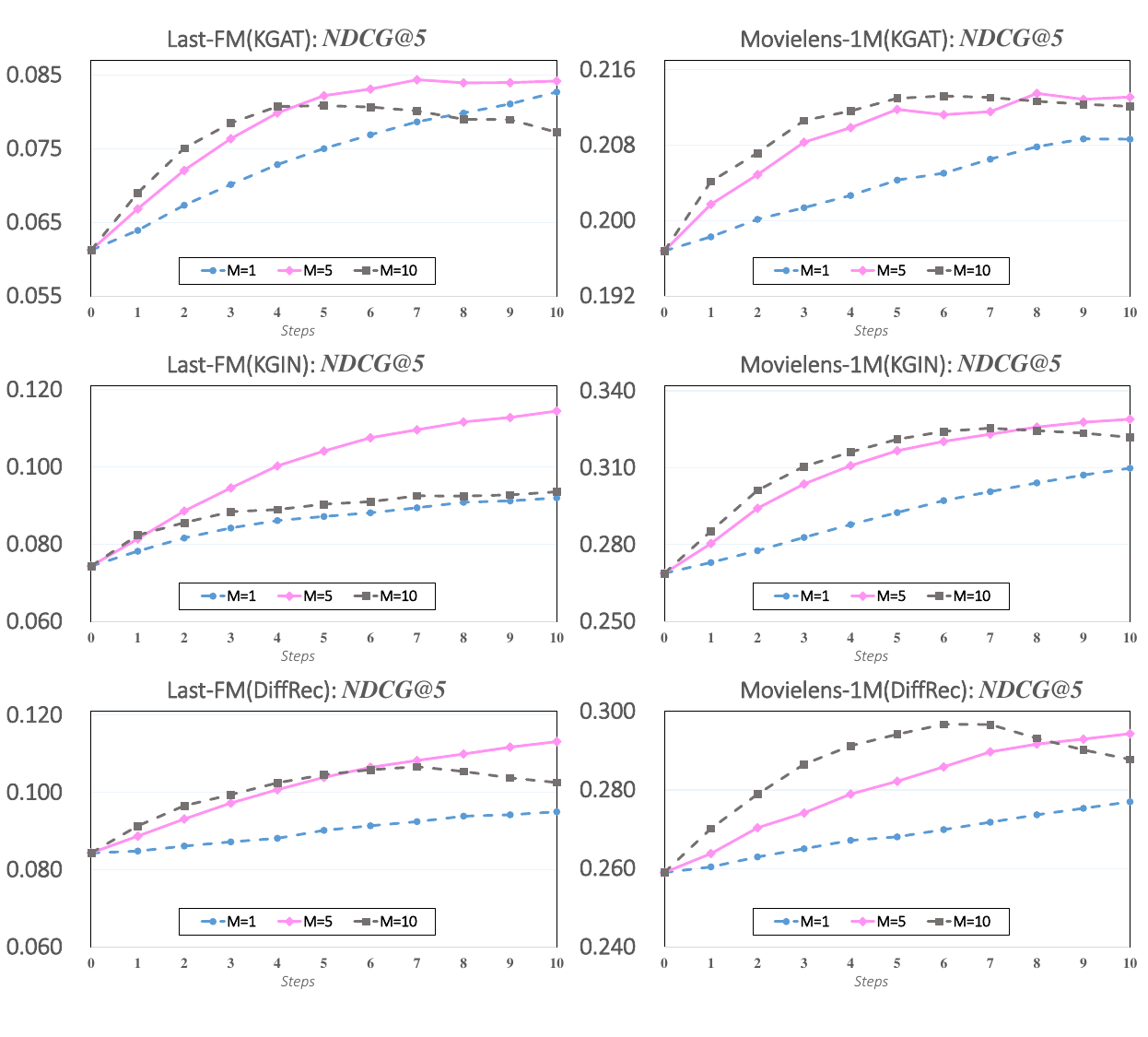}
		\subcaption{Impact of hyperparametric $M$.}
		\label{Figure7a}
	\end{minipage}
  \begin{minipage}[t]{0.32\textwidth}
		\includegraphics[width=0.95\textwidth]{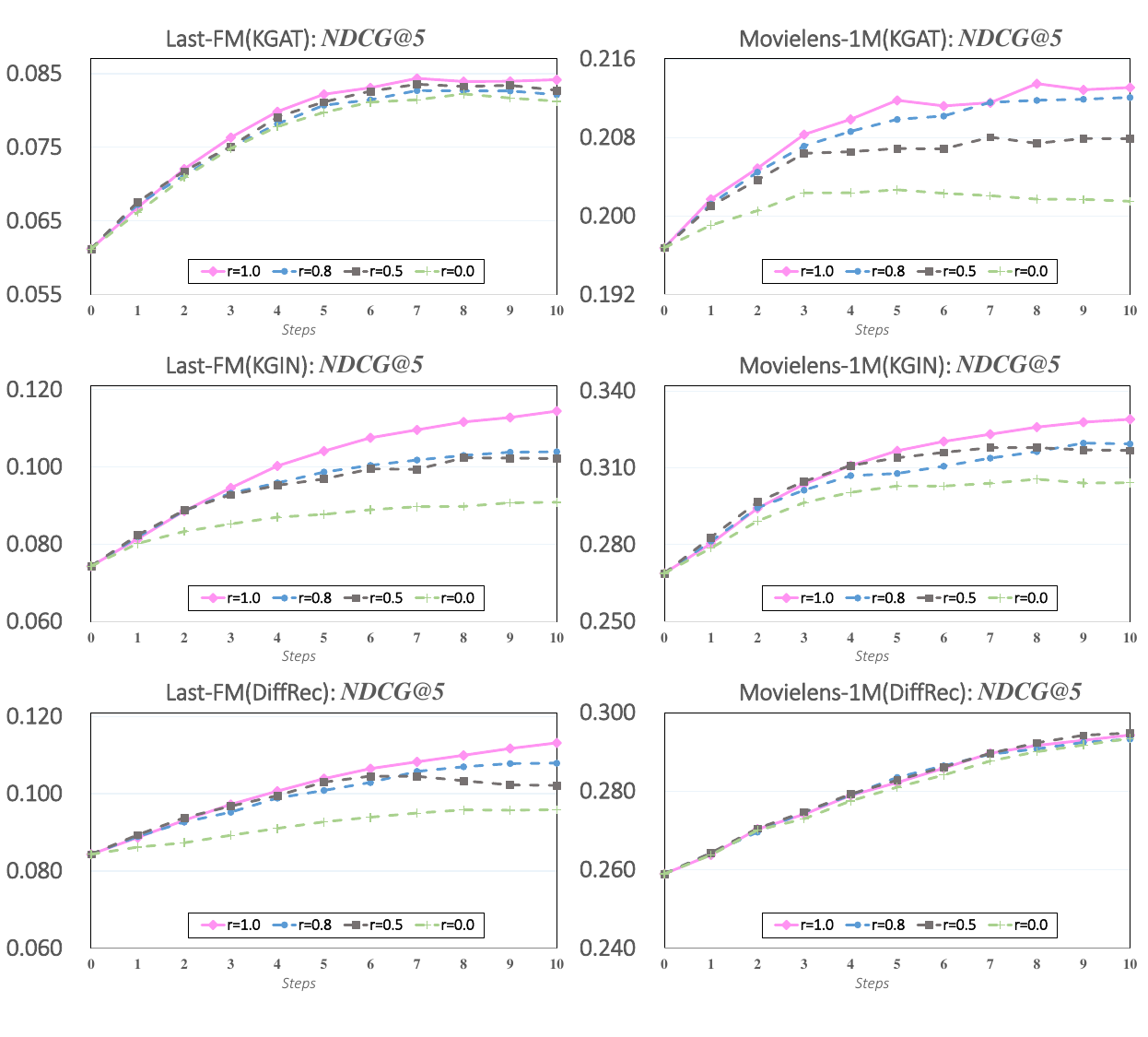}
		\subcaption{Impact of hyperparametric $r$.}
		\label{Figure7b}
	\end{minipage}
  \begin{minipage}[t]{0.32\textwidth} 
		\includegraphics[width=0.95\textwidth]{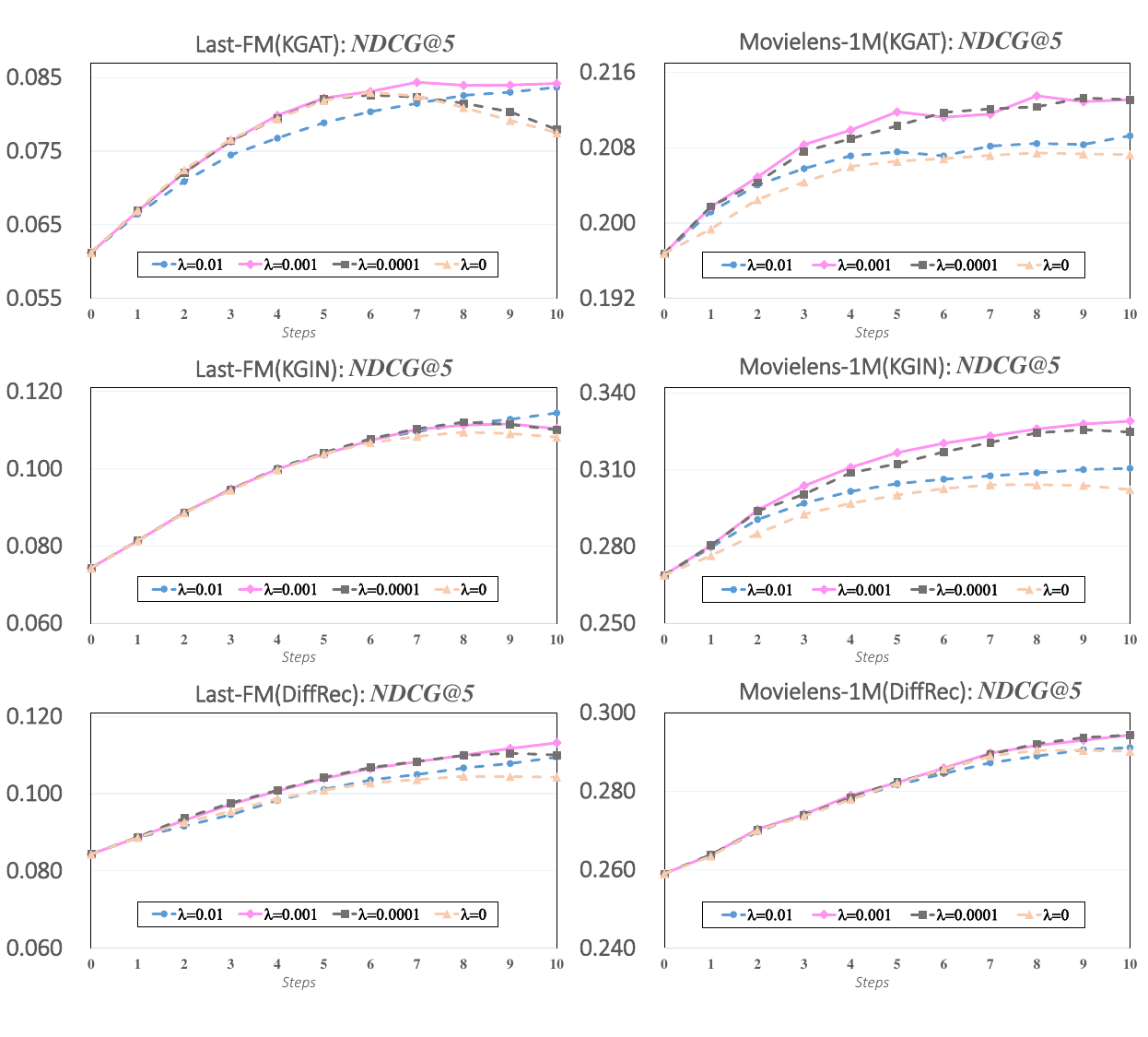}
		\subcaption{Impact of hyperparametric ${\lambda _\Omega }$.}
		\label{Figure7c}
  
	\end{minipage}
  \caption{Influence of different hyperparameters}

\end{figure*}

\subsubsection{Impact of $M$}
The quality of sampling representative items proxy for keyphrases is crucial. 
The goal is to explore an adequate sampling size nearest to conveying the semantics of keyphrase. 
We conducted an exploration of sampling quantities in $M=\{1, 5, 10\}$. As depicted in Figure 7(a), a few samples $(M=1)$ cannot 
fully capture the characteristics of keyphrases. Conversely, too many samples $(M=10)$ will result in a downward trend 
toward the tail of multistep critiquing. Choosing $M=5$ was based on the fact that despite a slightly weaker performance 
in the first few rounds, it maintained its ability well all over.
\subsubsection{Impact of $r$}
In this section, we explore the impact of higher-order items on surrogate keyphrases. Since the original model 
utilized considerable higher-order information during training phrase to enhance the keyphrase embedding, 
it is worthwhile to investigate how many higher-order items need to be sampled. We selected the proportion of 
one-hop items directly connected with keyphrases in $r=\{1.0, 0.8, 0.5, 0.0\}$. As illustrated in Figure 7(b), extracting 
more higher-order items in the preliminary usually yields greater improvements. However, later on, fatigue will occur, 
causing decreased benefits.
\subsubsection{Impact of ${\lambda _\Omega }$}
The strength of the regularizer has a great influence on the resistance of IPGC against forgetting. 
In Figure 7(c), while higher ${\lambda _\Omega }$ make the curve very smooth, but also greatly reduces its ability. 
And if ${\lambda _\Omega }$ is too small there will be some volatility in the critiquing process. Therefore, 
we operate for ${\lambda _\Omega }=0.001$, ensuring that the regularizer effective without compromising IPGC's adaptability.

\begin{figure}[]
  \centering
  \includegraphics[width=.9\linewidth]{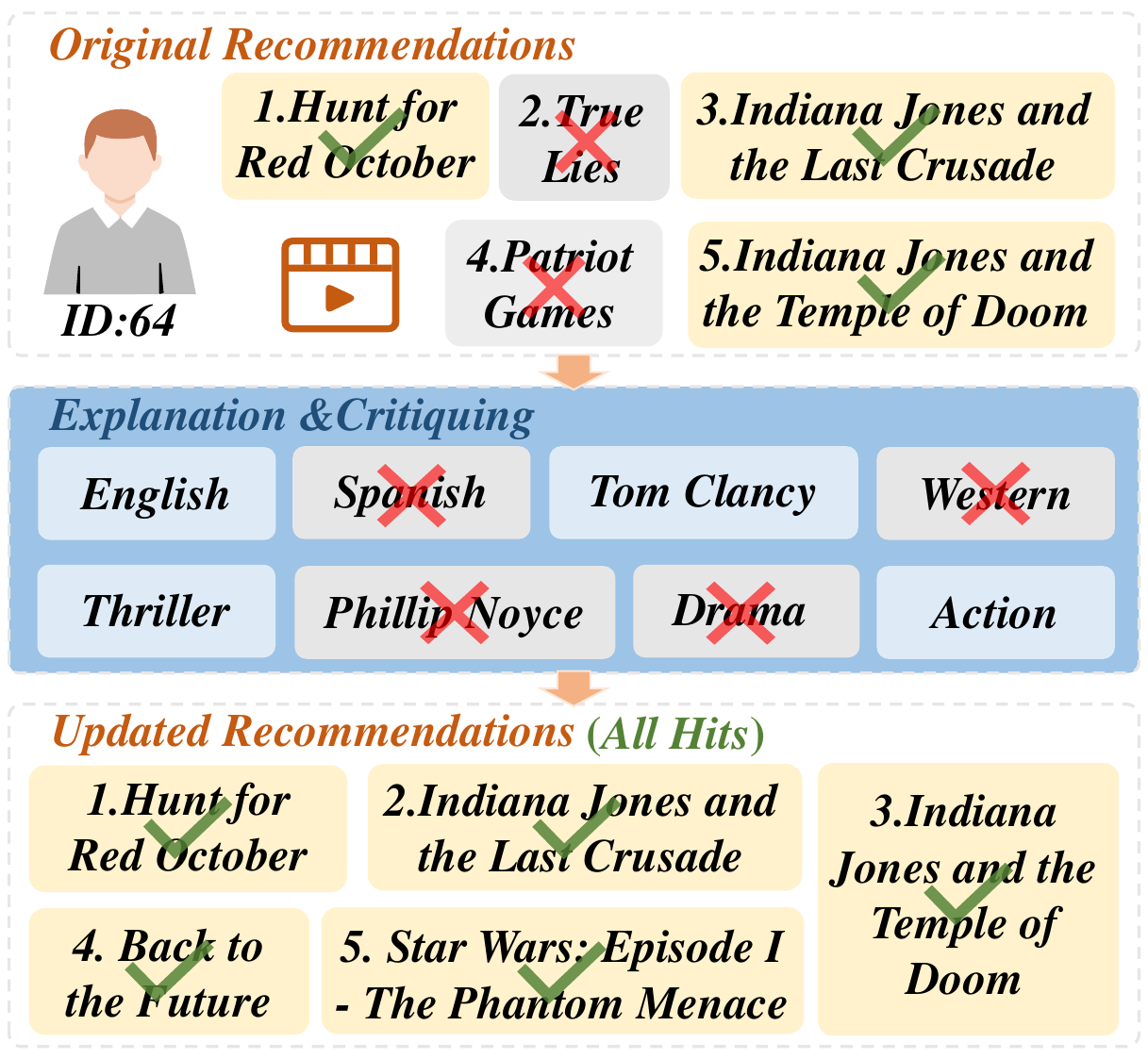}
  \caption{Example of a user ID 64 in Movielens.}
  \label{Figure8}
\end{figure}

\subsection{Case Study (RQ3)}
We illustrate the critiquing using a real case from the Movielens-1M. We randomly selected a user with his ID is 64 and 
demonstrated the simulation of his critiquing process. As seen in Figure 8, he disliked the 2nd and 4th movies 
in the original recommendation. Therefore, based on the explanatory keyphrases provided by the application, 
he made four valid negative critiquing: Spanish, Drama, Westerns, and Phillip Noyce. Based on these, IPGC update 
Top 5 recommendations with two films that users genuinely enjoyed while retaining the previous hits. This example 
illustrates that our IPGC framework can precisely refine recommendations according to user critiquing on the basis of 
the original model that is equipped with a high degree of accuracy.

\section{Conclusion and Future Work}
In this work, we introduced IPGC, a novel and versatile plugin designed to work with mostly CF models upon KG, providing a new option 
for refining recommendations using critiquing in list-based recommender scenarios. Experimental results demonstrate the 
effectiveness of the IPGC under various types of critiquing. It can be implemented in different architectures of 
KG recommender models and effectively alleviates catastrophic forgetting. Although IPGC still has some shortcomings, such 
as the necessity to work dependent on KG and the anti-forgotten regularization is not sota, the items proxy mechanism 
can efficiently utilize critiquing information to refine recommendations. For future work, 
we plan to (1) develop a more sophisticated proxy selection mechanism and (2) design a more powerful regularizer to enhance 
the performance of the critiquing plugin.

\section*{Acknowledgements}
This work is supported by National Natural Science Foundation of China under 62277028.




\clearpage 






\bibliographystyle{cas-model2-names}
\balance
\bibliography{refe}



\end{document}